\documentclass[twocolumn,superscriptaddress,showpacs,preprintnumbers,amsmath,amssymb]{revtex4}
\usepackage{amsmath}
\usepackage{amssymb}
\usepackage{dcolumn}
\usepackage{graphicx}
\usepackage{bm}

\begin{document}
\title{Computational studies of the glass-forming ability of model bulk 
metallic glasses}
\author{Kai Zhang}
\affiliation{Department of Mechanical Engineering and Materials Science, Yale University, New Haven, Connecticut, 06520, USA}
\author{Minglei Wang}
\affiliation{Department of Mechanical Engineering and Materials Science, Yale University, New Haven, Connecticut, 06520, USA}
\author{Stefanos Papanikolaou}
\affiliation{Department of Mechanical Engineering and Materials Science, Yale University, New Haven, Connecticut, 06520, USA}
\affiliation{Department of Physics, Yale University, New Haven, Connecticut, 06520, USA}
\author{Yanhui Liu}
\affiliation{Department of Mechanical Engineering and Materials Science, Yale University, New Haven, Connecticut, 06520, USA}
\author{Jan Schroers}
\affiliation{Department of Mechanical Engineering and Materials Science, Yale University, New Haven, Connecticut, 06520, USA}
\author{Mark D. Shattuck}
\affiliation{Department of Physics and Benjamin Levich Institute, The City College of the City University of New York, New York, New York, 10031, USA}
\affiliation{Department of Mechanical Engineering and Materials Science, Yale University, New Haven, Connecticut, 06520, USA}
\author{Corey O'Hern}
\affiliation{Department of Mechanical Engineering and Materials Science, Yale University, New Haven, Connecticut, 06520, USA}
\affiliation{Department of Physics, Yale University, New Haven, Connecticut, 06520, USA}
\affiliation{Department of Applied Physics, Yale University, New Haven, Connecticut, 06520, USA}

\date{\today}

\begin{abstract}
Bulk metallic glasses (BMGs) are produced by rapidly thermally
quenching supercooled liquid metal alloys below the glass transition
temperature at rates much faster than the critical cooling rate $R_c$
below which crystallization occurs.  The glass-forming ability 
of BMGs increases with decreasing $R_c$, and thus good glass-formers
possess small values of $R_c$.  We perform molecular dynamics
simulations of binary Lennard-Jones (LJ) mixtures to quantify how key
parameters, such as the stoichiometry, particle size difference,
attraction strength, and heat of mixing, influence the
glass-formability of model BMGs.  For binary LJ mixtures, we find that
the best glass-forming mixtures possess atomic size ratios (small to
large) less than $0.92$ and stoichiometries near $50$:$50$ by
number. In addition, weaker attractive interactions between the
smaller atoms facilitate glass formation, whereas negative heats of
mixing (in the experimentally relevant regime) do not change $R_c$
significantly. These studies represent a first step in the development
of computational methods for quantitatively predicting
glass-formability.
\end{abstract}

 \maketitle

\section{Introduction}

When supercooled liquids are rapidly quenched at rates $R$ exceeding a
critical value $R_c$, crystallization is avoided, and systems form
disordered solids such as bulk metallic glasses (BMGs). BMGs possess
high mechanical strength and can be processed so that they display
plastic~\cite{jan}, not brittle, response to applied deformations, which makes
them desirable materials for a variety of industrial and engineering
applications~\cite{inoue:2000}.  Avoiding crystallization in pure
metals requires enormously large cooling rates in excess of $10^{12}$
K/s.  However, bulk metallic glass-forming alloys have been developed
for which the critical cooling rate is more than nine orders of
magnitude lower, in the range $1 < R_c < 10^3$ K/s.  Understanding the
important physical quantities that determine the glass-forming ability
of multi-component alloys will allow us to develop even stronger
and less costly bulk metallic glasses.

\begin{figure*}
\includegraphics[width=2.32in]{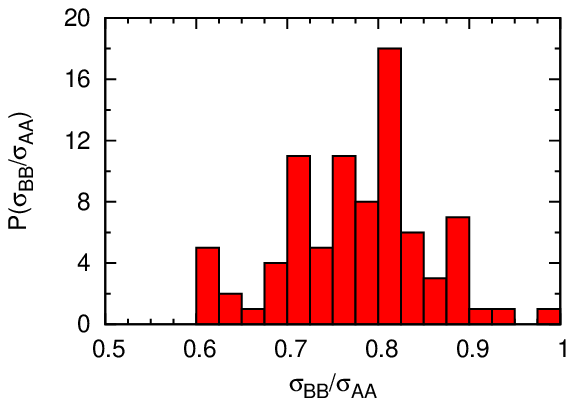}
\includegraphics[width=2.32in]{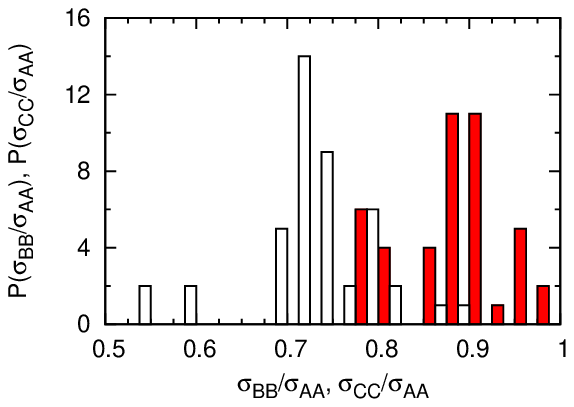}
\includegraphics[width=2.32in]{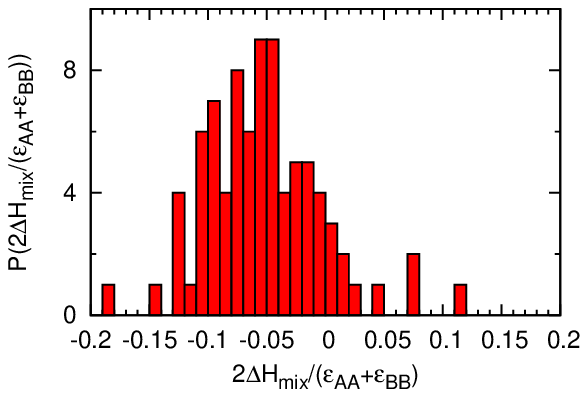}
\vspace{-0.3in}
\caption{(color online) (left) Probability distribution
$P(\sigma_{BB}/\sigma_{AA})$ of atomic size ratios (with $\sigma_{BB}< \sigma_{AA}$) in binary bulk metallic
glasses~\cite{miracle:2003}. (middle) Probability distributions of
atomic size ratios $P(\sigma_{BB}/\sigma_{AA})$ (shaded) and
$P(\sigma_{CC}/\sigma_{AA})$ (white) (with $\sigma_{CC} < \sigma_{BB}
< \sigma_{AA}$) in ternary BMGs~\cite{miracle:2003}. (right)
Probability distribution of the heats of mixing $\Delta H_{\rm mix}$
relative to the average cohesive energy $(\epsilon_{AA} +\epsilon_{BB})/2$ in binary BMGs~\cite{miracle:2003,takeuchi:2005}.}
\label{fig:experiment}
\end{figure*}

Prior research suggests that multi-component metallic alloys with
$T_g/T_m \gtrsim 0.67$ form BMGs, where $T_g$ and $T_m$ are the glass
transition and melting temperature,
respectively~\cite{turnbull:1969}.  In addition,
Inoue~\cite{inoue:2000} has emphasized three guidelines for enabling
BMG formation, rather than crystallization: 1) atomic size ratios
(small relative to large) of $\alpha < 0.89$ for at least two
constituents of the alloy; 2) large negative heats of
mixing~\cite{takeuchi:2005}; and 3) several atomic components. In
Fig.~\ref{fig:experiment}, we show the distributions of the atomic
size ratios and heats of mixing for common binary and ternary bulk
metallic glass-forming alloys~\cite{miracle:2003}. For binary systems,
the most probable atomic size ratio is $\alpha \approx 0.8$ and heat
of mixing is negative and roughly $6$-$7\%$ of the average cohesive
energy.

However, beyond these heuristic guidelines, there is no quantitative
and predictive understanding of the glass-forming ability in
multi-component alloys. (Note that there have been previous measurements of the
critical cooling rate in binary hard-sphere
systems~\cite{jalali:2004,jalali:2005}.) For model BMG-forming systems
with attractive interactions, we do not know the dependence of the
critical cooling rate on the stoichiometry, size ratios, and heats of
mixing of the constituent atomic species. For example, can
multi-component systems with large negative heats of mixing, but
smaller atomic size mismatches possess the same glass-forming ability
as systems with small negative heats of mixing but larger atomic size
mismatches?

We perform molecular dynamics simulations of model glass-forming
systems, binary Lennard-Jones mixtures of spherical particles, to
measure the critical cooling rate as a function of the size ratio,
number fraction, and interaction energy of the two particle species.
We find several important results. First, the critical cooling rate
decreases exponentially with the particle size ratio, $R_c \sim
\exp[-C(1-\alpha)^3]$, where $C$ depends on the number fraction of
small and large particles.  At a given size ratio $\alpha<1$, the
minimum critical cooling rate occurs at the number fraction
corresponding to equal volumes of the large and small particles.  In
addition, we find that at fixed number fraction and size ratio, the
critical cooling rate decreases strongly with decreasing cohesive
energy ratio of the small particles relative to the large ones,
$\epsilon_{BB}/\epsilon_{AA}$.  In contrast, variations of the heat of
mixing of the two species in the experimentally accessible range do
not affect $R_c$ significantly.  Thus, we have quantified several
design principles for improving glass formation in binary mixtures.

\section{Simulation methods}
\label{methods}

We perform constant number, volume, and temperature (NVT) molecular dynamics
(MD) simulations of binary Lennard-Jones (LJ) mixtures of $N=N_A+N_B$
spherical particles with the same mass $m$, but different diameters
$\sigma_{AA}$ and $\sigma_{BB}$, in periodic cubic cells with volume
$V=L^3$.  The particles interact pairwise via the LJ potential
\begin{equation}
\label{lj}
u(r_{ij}) = 4 \epsilon_{ij} \left[ \left(\frac{\sigma_{ij}}{r_{ij}}\right)^{12} - \left(\frac{\sigma_{ij}}{r_{ij}}\right)^6 \right],
\end{equation}
where $i,j \in \{A,B\}$, $B$ indicates the smaller particle,
$\sigma_{ij} = (\sigma_{ii}+\sigma_{jj})/2$ unless otherwise specified, and $\epsilon_{AA}$ and
$\epsilon_{BB}$ represent the cohesive energies for the $A$ and $B$
particles, respectively.  We quantify the heat of mixing using $\Delta
H_{\rm mix} = (\epsilon_{AA}+\epsilon_{BB})/2 - \epsilon_{AB}$. We employ the
shifted-force version of the LJ potential (Eq.~\ref{lj}) so that the
pair potential and force vanish for separations beyond the cutoff
distance $r_{\rm{cut}} = 3.5 \sigma_{ij}$~\cite{allen:1987}.
Energies, lengths, timescales, and temperatures are given in units of
$\epsilon_{AA}$, $\sigma_{AA}$, $\sigma_{AA}\sqrt{m/\epsilon_{AA}}$,
and $\epsilon_{AA}/k_B$, respectively, where the Boltzmann constant
$k_B$ is set to unity.

We study the glass-forming ability of binary LJ mixtures at fixed
packing fraction $\phi = N\sigma_{AA}^3(1+f_B(\alpha^3-1))\pi/6V =
0.5236$ as a function of the number fraction $f_B = N_B/N$, particle
size ratio $\alpha = \sigma_{BB}/\sigma_{AA}$, relative cohesive
energy $\epsilon_{BB}/\epsilon_{AA}$, and heat of mixing $\Delta
H_{\rm mix}$.  We only show results for $0.92 \le \alpha \le 1$ for
which solid solutions with FCC crystal structures are the equilibrium
phase ~\cite{hopkins:2011}. We initialize the systems at high
temperature $T_0=2.0$, using the Nos\'{e}-Hoover
thermostat~\cite{nose:1984,hoover:1985}, and then thermally quench the
systems exponentially, $T(t) = T_0 e^{- R t}$, from $T_0$ to $T_f =
10^{-2}$ at various rates $R$ over four orders of magnitude. (In
Appendix~\ref{appendixa}, we show that our results are not sensitive 
to the choice of the thermostat and the form of the cooling schedule.)  

Following the thermal quenches to $T_f$, we characterize the
structural properties of the system by measuring several quantities:
1) the local and global bond orientational order
parameters~\cite{steinhardt:1983,schreck:2010,{wang:2005}}
\begin{equation}
\label{local}
Q_6^l = \left( \frac{4\pi}{13} \sum_{m=-6}^{6} \frac{1}{N} \sum_{i=1}^N \frac{1}{n_i} \left| \sum_{j=1}^{n_i} Y_6^m(\theta _{ij},\phi _{ij})\right|^2 \right)^{1/2}
\end{equation}
\begin{equation}
\label{global}
Q_6^g = \left( \frac{4\pi}{13} \sum_{m=-6}^{6} \left| \frac{1}{N} \sum_{i=1}^N \frac{1}{n_i} \sum_{j=1}^{n_i} Y_6^m(\theta _{ij},\phi _{ij})\right|^2 \right)^{1/2},
\end{equation}
where $\theta_{ij}$ and $\phi_{ij}$ are the axial and polar angles
between each particle $i$ and its neighbors $j$, $Y_6^m$ are spherical
harmonics of degree $6$ and order $m$, and $n_i$ is the number of
nearest neighbors of particle $i$ within a cutoff distance of
$1.5\sigma_{ij}$; 2) local bond orientational order position correlation
function
\begin{equation}
\label{g6}
G_6(r) = \frac{4\pi}{13} \sum_{m=-6}^6\frac{ \left|  \sum_i \sum_{j \neq i} q_{6m}(\vec{r}_i)q_{6m}(\vec{r}_j) \delta (\vec{r} - \vec{r}_{ij}) \right|}{g(r)},
\end{equation}
where $g(r)= \sum_i \sum_{j \neq i} \delta (\vec{r} -
\vec{r}_{ij})$ is the radial distribution function and
$q_{6m}(\vec{r}_i) =n_i^{-1} \sum_{j=1}^{n_i} Y_6^m(\theta
_{ij},\phi _{ij})$; and 3) the crystal domain size.  These structural
quantities are averaged over at least $96$ independent quenching
trajectories. (In Appendix~\ref{appendixb}, we compare the results using these structural quantities.)  We consider system sizes from $N=500$ to $8788$
particles.

\begin{figure*}
\includegraphics[width=3.5 in]{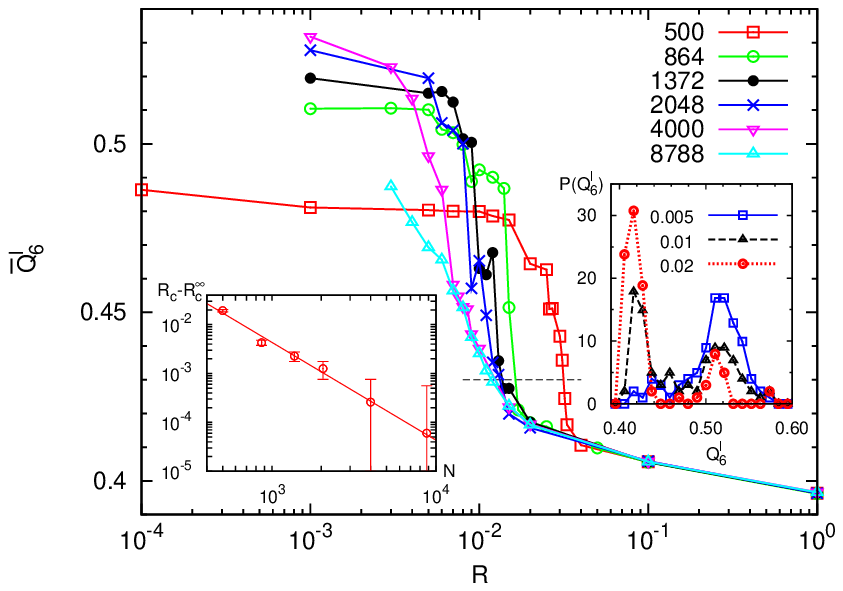}
\includegraphics[width=3.5 in]{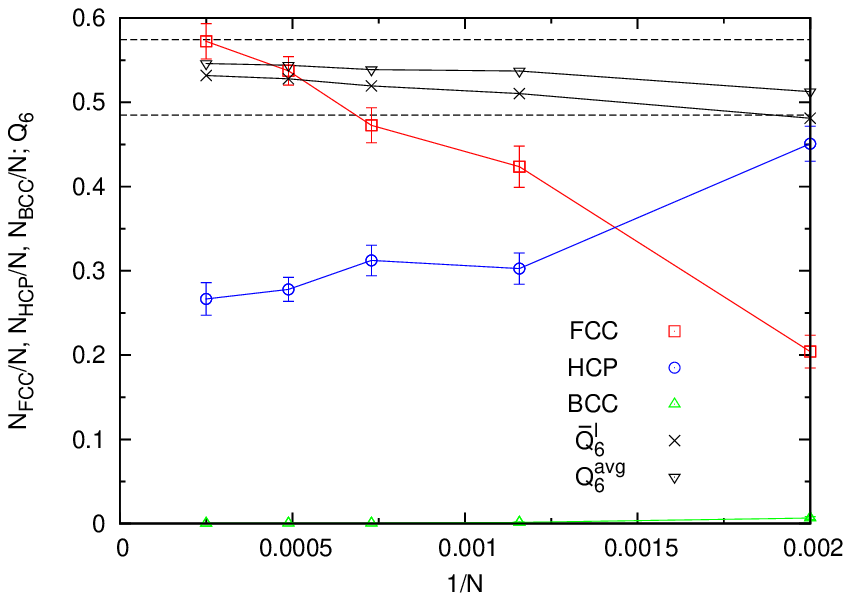}
\caption{(color online) (left) Median local bond orientational order
parameter ${\overline Q}_6^l$ for monodisperse Lennard-Jones (LJ)
systems following thermal quenches to $T_f = 0.01$ over a range of
cooling rates $R$ for system sizes $N = 500$, $864$, $1372$, $2048$,
$4000$, and $8788$. The critical cooling rate $R_c$ (defined, as discussed in the main text, as the
rate at which ${\overline Q}_6^l=0.43$ (dashed line)) approaches its
large-$N$ limit, $R_c^{\infty}$, as a power-law $R_c-R_c^{\infty} \sim
1/N^2$ (left inset).  (right inset) The probability distribution
$P(Q_6^l)$ for monodisperse LJ systems with $N=1372$ following
quenches to $T_f =0.01$ for cooling rates $R=0.02$ ($\circ$), $0.01$
($\triangle$), and $0.005$ ($\square$).  (right) Fraction of particles
that occur in HCP ($\circ$), FCC ($\square$), and BCC ($\triangle$)
crystal clusters as a function of $1/N$ for monodisperse LJ systems
following a quench to $T_f$ at cooling rate $R=10^{-3}<R_c$. At this
rate, the median local bond orientational order parameter ${\overline
Q}_6^l$ ($\times$) agrees with the value ($\triangledown$) obtained by
averaging $Q_6^l=0.575$ for FCC and $Q_6^l=0.485$ for HCP (dashed
lines) weighted by the fraction of particles in FCC and HCP clusters
in each sample.}
\label{fig:Q6}
\end{figure*}

\section{Results}
\label{results}

In this section, we characterize the structural properties of
LJ systems thermally quenched to temperature $T_f$ as a function of
the cooling rate $R$.  In the right inset of the left panel of
Fig.~\ref{fig:Q6}, we show the distribution $P(Q_6^l)$ of the local
bond orientational order parameter $Q_6^l$ for monodisperse LJ systems
with $N=1372$ particles. For fast cooling rates, {\it e.g.} $R =0.02$,
most of the quenched systems are structurally disordered, and
$P(Q_6^l)$ possesses a strong peak at small $Q_6^l \sim 0.41$. In
contrast, for slow cooling rates, {\it e.g.} $R =0.005$, most of the
quenched systems are ordered, and $P(Q_6^l)$ possesses a strong peak
at a larger value of $Q_6^l \sim 0.51$.  For intermediate cooling
rates, the distribution $P(Q_6^l)$ becomes strongly bimodal, which
indicates that the systems possess disordered as well as ordered
regions. In the main panel of Fig.~\ref{fig:Q6} (left), we show the
median ${\overline Q}_6^l$ versus the logarithm of the cooling rate
$R$ for several system sizes.  For each system size, ${\overline
Q}_6^l$ first increases modestly with decreasing cooling rate,
followed by a rapid increase at intermediate rates, and then it
plateaus with further decreases.  We define the critical cooling rate,
$R_c$, as the rate at which the median local bond orientational order
parameter crosses the threshold value ${\overline Q}_6^l = Q_0 =
0.43$. We chose the threshold $Q_0$ for several reasons: 1) $Q_0$
captures the steep rise in ${\overline Q}_6^l$ with decreasing cooling
rate, 2) $Q_0$ is in the region of $Q_6^l$ between the two peaks in
$P(Q_6^l)$ that occur at intermediate cooling rates (right inset of
left panel of Fig.~\ref{fig:Q6}), and 3) $Q_0$ is a value for which
$Q_6^l(R)$ becomes system size independent for intermediate and fast
cooling rates.

\begin{figure}
\includegraphics[width=1.68 in]{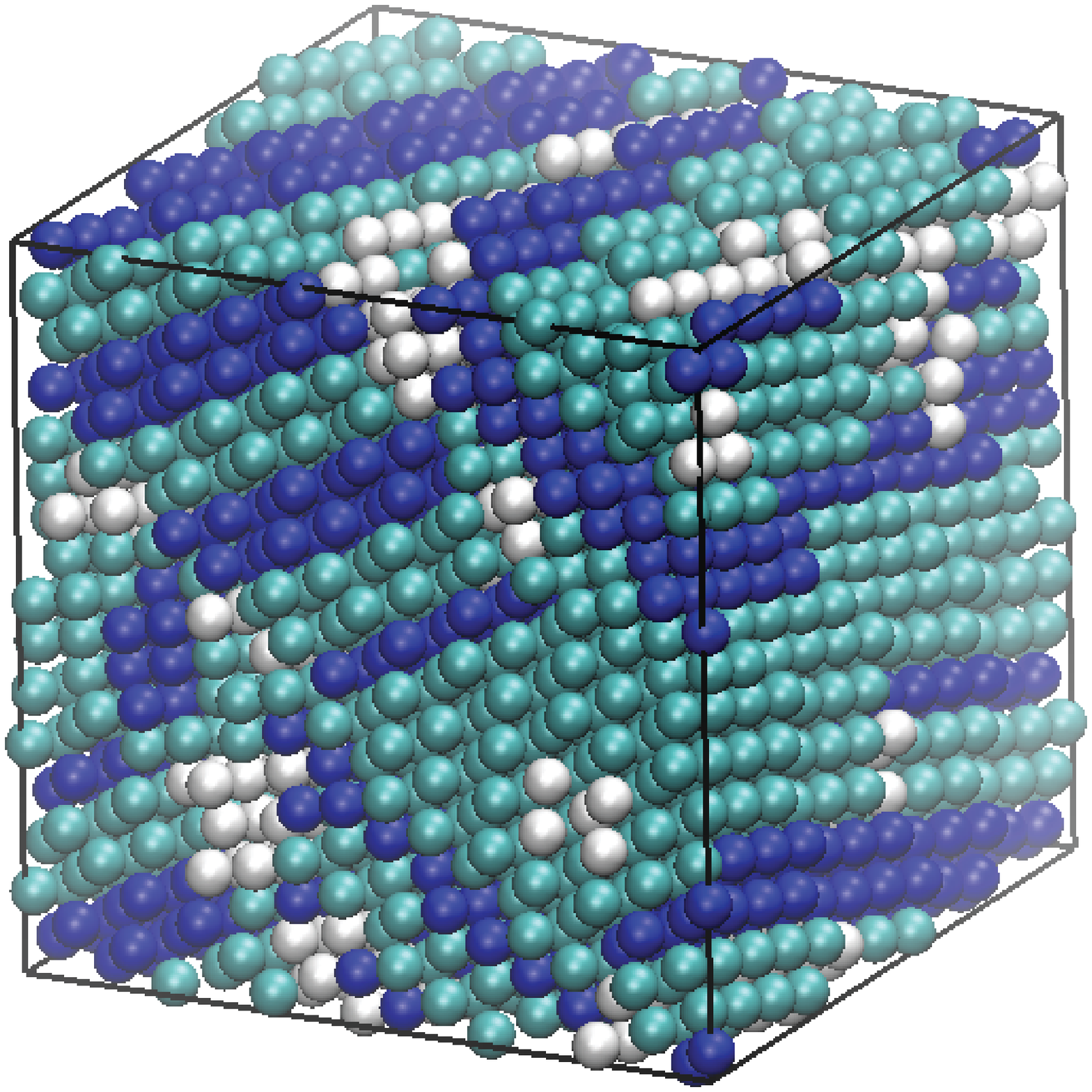}
\includegraphics[width=1.68 in]{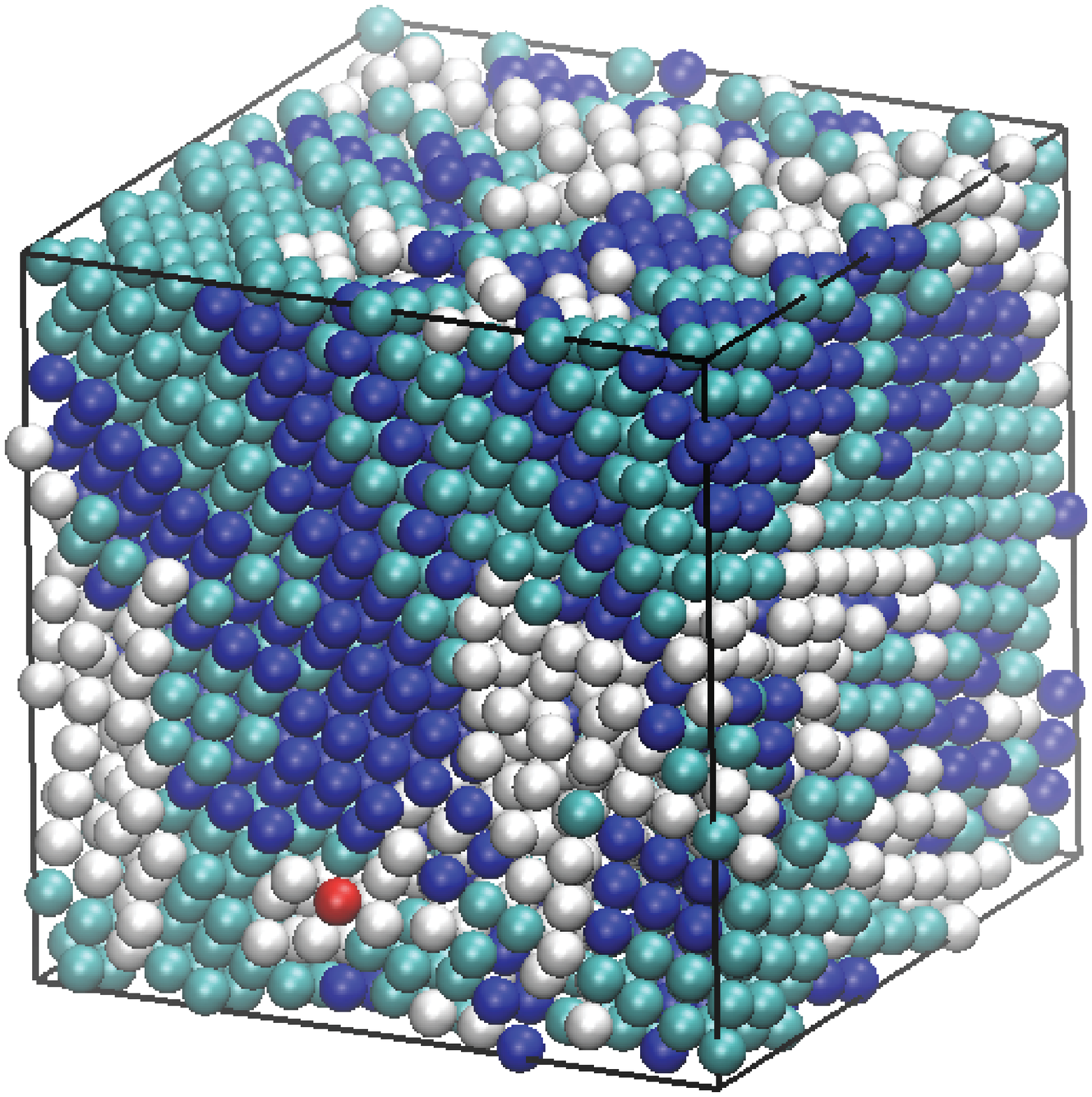}
\caption{(color online) Crystalline clusters obtained in monodisperse LJ
systems with $N=4000$ following thermal quenches to $T_f = 0.01$ at
cooling rates $R=10^{-3}$ (left) and $10^{-2}$ (right). Particles are
colored according to whether they belong to FCC (cyan), HCP (blue),
BCC (red), or non-crystalline (white) domains.}
\label{fig:dxasnapshot}
\end{figure}

Note that the distribution of the global bond orientational order
parameter $P(Q_6^g)$ also becomes bimodal and the median ${\overline
Q}_6^g$ increases rapidly with decreasing cooling rate.  (See
Appendix~B.) However, the global bond orientational order parameter
quantifies crystallization of the {\it entire} system, which is
influenced more by the slow dynamics of crystal growth, rather than
the initial nucleation of crystalline domains.

\begin{figure*}
\includegraphics[width=3.5in]{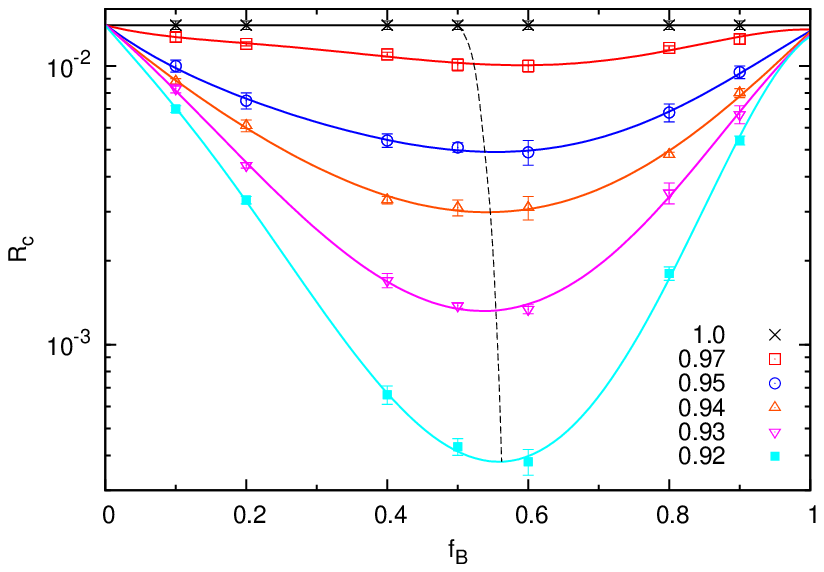}
\includegraphics[width=3.5in]{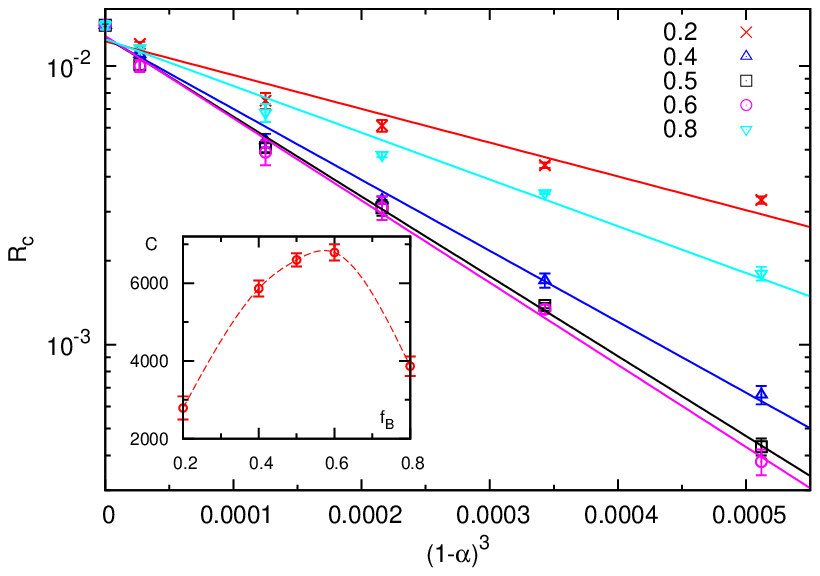}
\caption{(color online) (left) Critical cooling rate $R_c$ for binary
LJ mixtures with $N=1372$ as a function of the number fraction $f_B$
for several particle size ratios $\alpha=1.0$, $0.97$, $0.95$, $0.94$,
$0.93$, and $0.92$. The solid lines are sixth-order least-square fits
to the data for $R_c$. The dashed line connects the number fractions $f_B^*
=1/(1+\alpha^3)$ at which the $A$ and $B$ particles occupy the same
volume. (right) $R_c$ versus $(1-\alpha)^3$ for binary LJ mixtures
with $f_B= 0.2$, $0.4$, $0.5$, $0.6$, and $0.8$. The error bars for
$R_c$ are determined by the cooling rate increment $\Delta R =
10^{-3}$. The inset shows the coefficient $C(f_B)$ of the exponential
decay of $R_c\sim \exp[ -C(1-\alpha)^3]$.}
\label{fig:rc}
\end{figure*}

The value of the bond orientational order parameter depends on the
crystal structure that forms during the thermal quenching process.
Thus, we employed a crystal analysis algorithm to identify the
crystalline clusters (FCC, HCP~\cite{stukowski:2012}, or BCC) for
cooling rates $R \lesssim R_c$.  For example, $Q^l_6 \approx 0.575$
for an ideal face-centered cubic (FCC) structure, whereas it is
$\approx 0.485$ for an ideal hexagonal close packed (HCP) structure.
This difference explains the increase in ${\overline Q}_6^l$ for $R
\ll R_c$ as $N$ increases in the main panel of Fig.~\ref{fig:Q6}
(left).  In Fig.~\ref{fig:Q6} (right), we show that small systems $N
\le 500$ mainly crystallize to HCP structures~\cite{footnote:1}, while
larger systems crystallize predominantly to FCC structures. For low
cooling rates, the median local bond orientational order parameter
${\overline Q}_6^l$ can be obtained by averaging the $Q_6^l$ values
for FCC and HCP structures weighted by the fraction of particles in
FCC and HCP clusters in each sample. (See Fig.~\ref{fig:Q6} (right).)
We show snapshots of the thermally quenched structures for
monodisperse LJ systems using two cooling rates in
Fig.~\ref{fig:dxasnapshot} with FCC, BCC, HCP, and non-crystalline
regions shaded different colors.

We show the system-size dependence of the critical cooling rate $R_c$
for monodisperse LJ systems in the left inset to the left panel of
Fig.~\ref{fig:Q6}. We find that $R_c$ decreases with increasing system
size and approaches its large-$N$ limit, $R_c^{\infty} \approx 0.01$,
as a power law $R_c - R_c^{\infty} \sim 1/N^2$.  It is interesting
that the approach to $R_c^{\infty}$ scales as $1/N^2$, which is 
faster than the $1/N$ scaling typical for first-order transitions. In
contrast to hard-sphere systems~\cite{rintoul:1996}, crystallization
in monodisperse LJ systems is more difficult at large $N$.  In small
monodisperse LJ systems ($N \le 500$), the critical nucleus is
sufficiently large that it interacts with its periodic
images~\cite{honeycutt:1984,honeycutt:1986}, which reduces the
interfacial energy of crystal nuclei and enhances the formation of
single crystals.

We now focus on binary LJ systems at fixed $N=1372$ and cohesive
energy ratio $\epsilon_{BB}/\epsilon_{AA}=1$ and study the
glass-forming ability as a function of the size ratio $\alpha$ and
number fraction $f_B$. For $\alpha \lesssim 1$, the smallest
$R_c(\alpha, f_B)$ ({\it i.e.} best glass-former) is obtained in
systems with approximately equal numbers of $A$ and $B$ particles,
$f_B^* \approx 0.5$, as shown in Fig.~\ref{fig:rc} (left).  As
$\alpha$ decreases, the minimum in $R_c(\alpha, f_B)$ deviates from
$f_B^* \approx 0.5$ and follows $f_B^*=1/(1+\alpha^3)$ for which the
$A$ and $B$ particles occupy the same volume (reaching $f_B^* \approx
0.56$ at $\alpha=0.92$).  As shown in Fig.~\ref{fig:rc} (right), at
each $f_B$, $R_{c}$ decreases exponentially with decreasing size
ratio, $R_c(\alpha, f_B) = R_c(1, f_B) \exp[-C(f_B)
(1-\alpha)^3]$. This result implies that $R_c$ drops from $10^{-2}$ to
$10^{-11}$--$10^{-25}$ for binary systems of composition
$f_B=0.2$--$0.8$ with size ratio $\alpha = 0.8$ (the most common size
ratio in binary bulk metallic glass formers), which is 9--23 orders
slower than the $R_c$ at $\alpha = 1$. We also note that for a given
cooling rate $R$, the glass-forming regime, {\it i.e.} the range of
number fractions for which $R > R_c$, expands with decreasing
$\alpha$.

\begin{figure}
\includegraphics[width=3.5in]{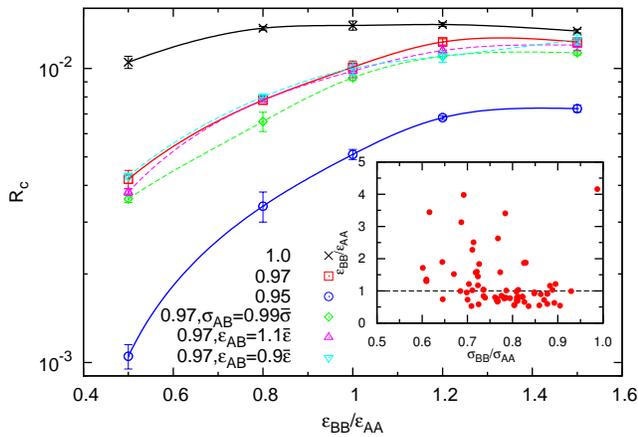}
\caption{(color online) Critical cooling rate $R_c$ as a function of
the cohesive energy ratio $\epsilon_{BB}/\epsilon_{AA}$ for binary LJ
mixtures with $N=1372$, number fraction $f_B=0.5$, and size ratios
$\alpha = 1.0$, $0.97$, and $0.95$. The solid lines indicate results
for the mixing rule $\sigma_{AB} =\bar{\sigma} \equiv (\sigma_{AA} + \sigma_{BB})/2$ and
$\epsilon_{AB} = \bar{\epsilon} \equiv (\epsilon_{AA} + \epsilon_{BB})/2$. The dashed lines
indicate results for positive ($\epsilon_{AB} = 0.9\bar{\epsilon}$) ($\triangledown$) and negative heats of mixing
($\epsilon_{AB} = 1.1\bar{\epsilon}$) ($\triangle$)
with $\sigma_{AB} = \bar{\sigma}$ and for bond
shortening $\sigma_{AB} = 0.99\bar{\sigma}$
with $\Delta H_{\rm mix}=0$ ($\diamond$). (inset) Cohesive energy ratio
$\epsilon_{BB}/\epsilon_{AA}$ versus the atomic size ratio
$\sigma_{BB}/\sigma_{AA}$ for common binary metallic glass
formers~\cite{miracle:2003}.}
\label{fig:rc_epsilon}
\end{figure}

For the results presented so far, we set the cohesive energy ratio
$\epsilon_{BB}/\epsilon_{AA} = 1$. However, as shown in the inset to
Fig.~\ref{fig:rc_epsilon}, the cohesive energy between like species is
different for the two components for most binary bulk metallic glass
formers.  In Fig.~\ref{fig:rc_epsilon}, we show $R_c$ as a function of
$\epsilon_{BB}/\epsilon_{AA}$ for binary LJ mixtures with $N=1372$ at
fixed $f_B = 0.5$, $(\epsilon_{AA}+\epsilon_{BB})/2=1$, and heat of
mixing $\Delta H_{\rm mix} = 0$, assuming $\Delta H_{\rm mix} =
(\epsilon_{AA}+\epsilon_{BB})/2 -
\epsilon_{AB}$~\cite{miedema:1975,boom:1976a,boom:1976b} and the
mixing rules $\epsilon_{AB} =(\epsilon_{AA}+\epsilon_{BB})/2$ and
$\sigma_{AB} = (\sigma_{AA}+\sigma_{BB})/2$. We find that the
glass-forming ability increases ({\it i.e.} $R_c$ decreases) as
$\epsilon_{BB}/\epsilon_{AA}$ decreases below $1$.  This result is
consistent with the fact that most binary glass formers with $0.8 <
\alpha < 1$ possess $\epsilon_{BB}/\epsilon_{AA} <
1$~\cite{miracle:2003}. (See the inset to Fig.~\ref{fig:rc_epsilon}.)

\begin{figure*}
\includegraphics[width=2.32 in]{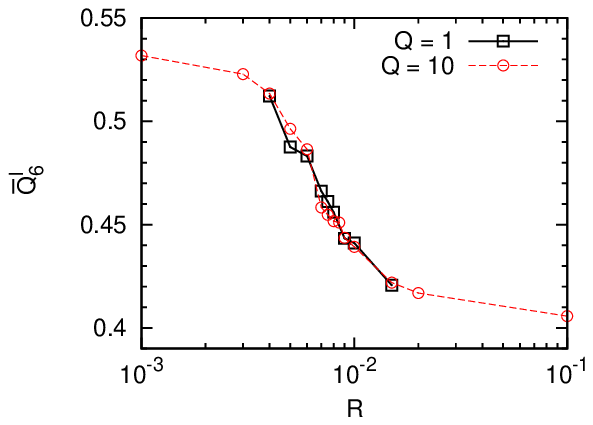}
\includegraphics[width=2.32 in]{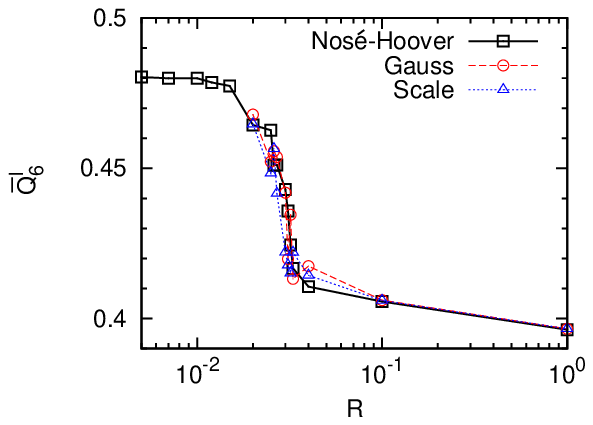}
\includegraphics[width=2.32 in]{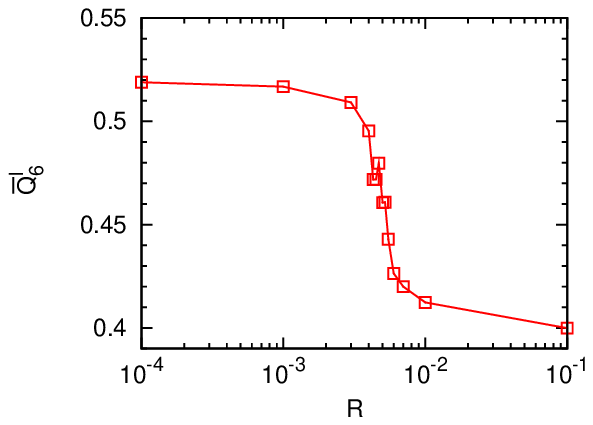}
\caption{(left) Median local bond-orientational order parameter
${\overline Q}_6^l$ versus the cooling rate $R$ for monodisperse LJ
systems with $N=4000$ using the Nos\'{e}-Hoover thermostat with
thermal inertia parameter $Q=1$ ($\boxdot$) and $10$ ($\odot$) in
units of $m\sigma_{AA}^2$. (middle) Median local bond-orientational
order parameter ${\overline Q}_6^l$ versus $R$ for monodisperse LJ
systems with $N=500$ using several thermostats: Nos\'{e}-Hoover
($\boxdot$), Gaussian constraint ($\odot$), and {\it ad hoc} velocity
rescaling $\vartriangle$. (right) Median local bond-orientational
order parameter ${\overline Q}_6^l$ versus the cooling rate $R$ for
monodisperse systems with $N=1372$ for a linear thermal quenching
protocol, $T(t) = T_0 - R t$.}
\label{fig:Q6method}
\end{figure*}

Inoue's guidelines~\cite{inoue:2000} suggest that a negative heat of
mixing $\Delta H_{\rm mix} < 0$ enhances the glass-forming ability of
BMGs.  The rationale is that a negative heat of mixing makes the mixed
and geometrically frustrated state energetically favorable compared to
the phase separated state.  Fig.~\ref{fig:experiment} (right)
shows that $\Delta H_{\rm mix}$ is approximately $5$-$10\%$ of the
average cohesive energy of the two components,
$(\epsilon_{AA}+\epsilon_{BB})/2$, for most binary
BMGs~\cite{varley:1954,miracle:2003,takeuchi:2005}. However, we show
in Fig.~\ref{fig:rc_epsilon} that binary LJ mixtures with heats of
mixing in the range $2 \Delta H_{\rm
mix}/(\epsilon_{AA}+\epsilon_{BB}) = \pm 0.1$ possess the same
critical cooling rate $R_c$ as those with $\Delta H_{\rm mix}=0$ over
the full range of size ratios studied.

Why then do most BMGs possess $\Delta H_{\rm mix} < 0$?  One
possibility is that negative heats of mixing are correlated with
strong bonding between atomic species, which can be modeled as bond
shortening ($\sigma_{AB} < (\sigma_{AA} +\sigma_{BB}
)/2$)~\cite{cheng:2009,liu:2009,senkov:2012}.  In
Fig.~\ref{fig:rc_epsilon}, we show that only a $1\%$ bond shortening,
$\sigma_{AB} = 0.99(\sigma_{AA} +\sigma_{BB})/2$, can give rise to a
finite decrease in the critical cooling rate $R_c$.

\section{Conclusion}

The glass formability of bulk metallic glass-forming alloys can be
characterized by the critical cooling rate $R_c$, below which the
system possesses crystalline domains.  The best bulk metallic glasses
are those with the lowest values for $R_c$.  However, the key
parameters that determine $R_c$ are not currently known, and thus BMGs
are mainly developed through a trial and error process. As a first
step in computational design of BMGs, we performed molecular dynamics
simulations of coarse-grained models for BMGs, binary Lennard-Jones
mixtures, and measured $R_c$ as a function of the number fraction,
size ratio, relative cohesive energy, and heat of mixing of the two
atomic species.  We measured the local bond orientational order
parameter to quantify the degree of crystallization that had occurred
in systems during thermal quenches from high to low temperature over
more than four orders of magnitude in the cooling rate.  It is known
that weakly polydisperse LJ systems are poor glass-formers; we
quantified this statement by showing that the critical cooling rate
decreases exponentially with increasing particle size ratio $\alpha$,
$R_c \sim \exp[-C(1-\alpha)^3]$.  Further, at a given size ratio
$\alpha<1$, the minimum critical cooling rate occurs at the number
fraction corresponding to equal volumes of the large and small
particles of equal mass.  In addition, we find that at fixed number
fraction and size ratio, the critical cooling rate decreases strongly
with decreasing cohesive energy ratio of the small particles relative
to the large ones, $\epsilon_{BB}/\epsilon_{AA}$.  This result may
explain why most experimentally obtained binary BMGs possess
$\epsilon_{BB}/\epsilon_{AA} < 1$. In contrast, variations of the heat
of mixing of the two species in the experimentally accessible range
(several per cent of the average cohesive energy) do not affect $R_c$
for binary LJ mixtures significantly.  However, bond shortening of
only several percent relative to $\sigma_{AB} =
(\sigma_{AA}+\sigma_{BB})/2$~\cite{cheng:2009,liu:2009,senkov:2012}
does give rise to significant changes in $R_c$.  Recent experiments
have suggested that negative heats of mixing are correlated with
bond-shortening, which may explain why most experimentally obtained
BMGs possess negative heats of mixing.  In future studies, we will
characterize the glass-forming ability and crystallization processes
in ternary and quaternary LJ mixtures using MD simulations, energy
minimization, and genetic algorithms.

\begin{acknowledgments}
We thank Frans Spaepen and Michael Falk for helpful discussions. The
authors acknowledge primary financial support from the NSF MRSEC
DMR-1119826 (KZ and MW) and partial support from NSF grant numbers 
DMR-1006537 (CO) and CBET-0968013 (MS).
\end{acknowledgments}

\begin{appendix}

\section{Thermostat and Quenching Protocol}
\label{appendixa}

In this appendix, we provide additional details of the molecular
dynamics (MD) simulations used to thermally quench Lennard-Jones (LJ) 
systems from high temperature liquids to low temperature glasses.  The
LJ liquids were first equilibrated at high temperature
$T_0=2.0$ using constant number $N$, volume $V$, and temperature $T$
MD simulations, and cooled exponentially $T(t) = T_0 e^{- R t}$ to low
temperature $T_f=10^{-2}$. The temperature was controlled using the
Nos\'{e}-Hoover thermostat~\cite{nose:1984,hoover:1985} with thermal
inertia parameter $Q=1$, and the equations of motion were integrated
using a Newton's method technique~\cite{frenkel:2002} with time step
$\Delta t = 10^{-3}$.  In Fig.~\ref{fig:Q6method} (left), we show for
monodisperse LJ systems with $N=4000$ that the dependence of the
median local bond orientational parameter ${\overline Q}_6^l$ on rate $R$
is the same for $Q=1$ and $10$.

We also investigated the extent to which the thermostat affects the
critical cooling rate, below which the systems crystallize. In
Fig.~\ref{fig:Q6method} (center), we show that ${\overline Q}_6^l$
versus $R$ is the same for monodisperse LJ systems with $N=500$ when
the temperature is controlled using the Nos\'{e}-Hoover, Gaussian
constraint, and {\it ad hoc} velocity rescaling
thermostats~\cite{brown:1984,allen:1987}. Thus, the choice of the
thermostat does not influence the measurement of $R_c$. We also varied
the form of the thermal quenching protocol. In Fig.~\ref{fig:Q6method}
(right), we show that a linear cooling schedule, $T(t) = T_0 - R t$,
gives qualitatively the same results for ${\overline Q}_6^l$ versus
$R$ as an exponential temperature ramp.

\begin{figure}
\includegraphics[width=3.5 in]{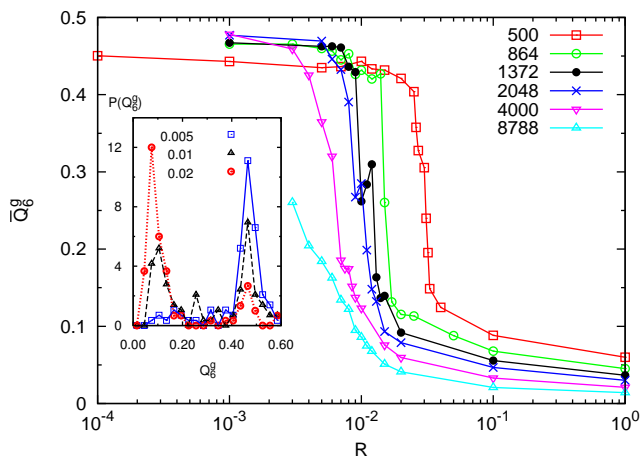}
\caption{(color online)  Median global bond orientational order
parameter ${\overline Q}_6^g$ for monodisperse LJ systems following
thermal quenches to $T_f = 0.01$ over a range of cooling rates $R$ for
system sizes $N = 500$, $864$, $1372$, $2048$,  $4000$, and $8788$. (inset) The probability distribution
$P(Q_6^g)$ for monodisperse LJ systems with $N=1372$ following
quenches to $T_f =0.01$ for cooling rates $R=0.02$ ($\circ$), $0.01$
($\triangle$), and $0.005$ ($\square$). }
\label{fig:Q6global}
\end{figure}

\begin{figure*}
\includegraphics[width=2.32 in]{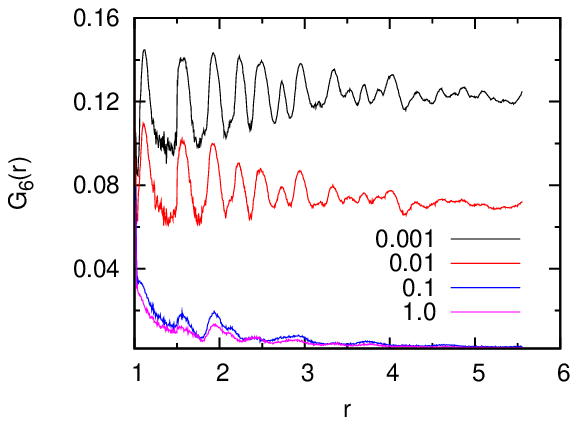}
\includegraphics[width=2.32 in]{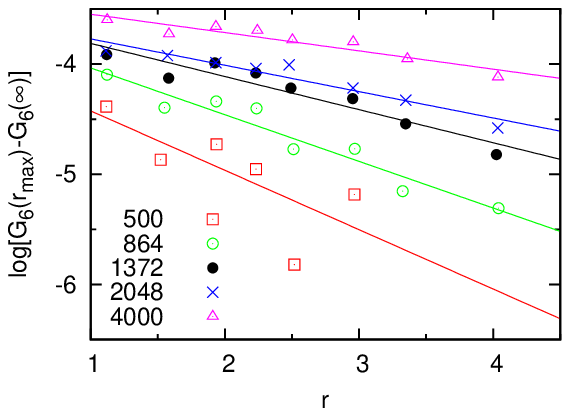}
\includegraphics[width=2.32 in]{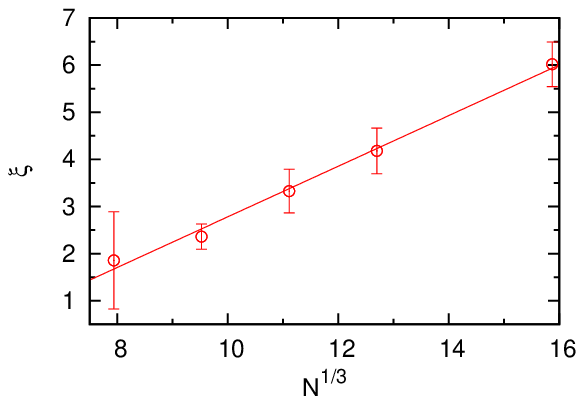}
\caption{(color online)  (left) Local bond orientational order correlation
function $G_6(r)$ for monodisperse LJ systems with $N=1372$ at several
cooling rates $R=1$, $10^{-1}$, $10^{-2}$, and $10^{-3}$. (middle) The
decay of the local maxima in $G_6(r)$ versus distance $r$ for monodisperse
LJ systems at cooling rate $R=10^{-3}$ for several system sizes
$N=500$, $864$, $1372$, $2048$, and $4000$. (right) Correlation length
$\xi$ from the decay of the local bond orientational order  
correlation function versus the linear dimension of the system 
$N^{1/3}$ for monodisperse LJ systems at cooling rate
$R=10^{-3}$. The solid line has slope $\approx 0.5$.}
\label{fig:G6}
\end{figure*}

\section{Characterization of Crystalline Order}
\label{appendixb}

In this Appendix, we describe several metrics (in addition to the
local bond orientational order parameter $Q_6^l$ in Eq.~\ref{local})
to characterize the degree of crystalline order of thermally quenched
LJ systems.  In contrast to $Q_6^l$, the global bond orientational
order parameter $Q_6^g$ in Eq.~\ref{global} quantifies the degree of
crystallization over the entire system. The median global bond
orientational order parameter ${\overline Q}_6^g$ versus cooling rate
$R$ for monodisperse LJ systems for several system sizes is shown in
Fig.~\ref{fig:Q6global}. ${\overline Q}_6^g$ shows a rapid increase
near the critical cooling rate $R_c$ as found for ${\overline Q}_6^l$.
However, $R_c$ (defined by a threshold such as ${\overline
Q}_6^g=0.3$) appears to decrease to zero in the large system limit.
This trend occurs because it takes an increasing amount of time (and
thus slower cooling rates) for crystal nuclei to grow and for the
system to reach the same ${\overline Q}_6^g$ as that obtained in
smaller systems.

In Fig.~\ref{fig:G6} (left), we show the local bond orientational
order correlation function (Eq.~\ref{g6}) for monodisperse
LJ systems with $N=1372$ for several cooling rates.  We find that
$G_6(r)$ plateaus at large $r$ and the plateau value $G_6(\infty)$
increases with decreasing cooling rate $R$.  For partially crystalline
systems, $G_6(r)$ decays to $1/\sqrt{N_d}$ at large distances, where
$N_d$ is the number of independent crystalline domains. For disordered
systems, $G_6(r)$ decays to $1/\sqrt{N_b}$, where $N_b$ is the total
number of nearest neighbor particles~\cite{schreck:2010}.  We find that the
deviation $G_6(r_{{\rm max}} )-G_6(\infty)$, where $G_6(r_{\rm max})$
are the local maxima in $G_6(r)$, decays exponentially $\sim
e^{-r/\xi}$ with correlation length $\xi$. (See Fig.~\ref{fig:G6}.) The
correlation length $\xi$ grows linearly with the linear size of the
system $N^{1/3}$ for cooling rates $R < R_c$.

\begin{figure*}
\includegraphics[width=3.5 in]{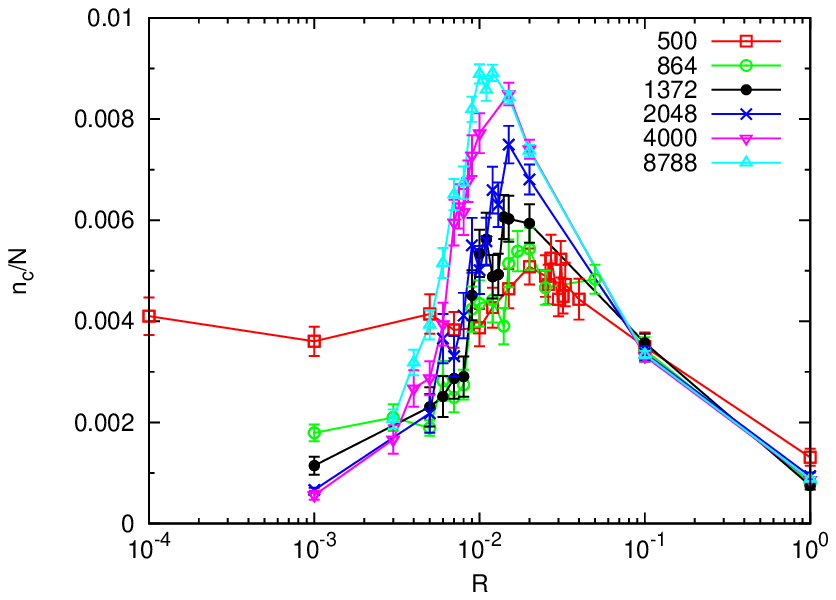}
\includegraphics[width=3.5 in]{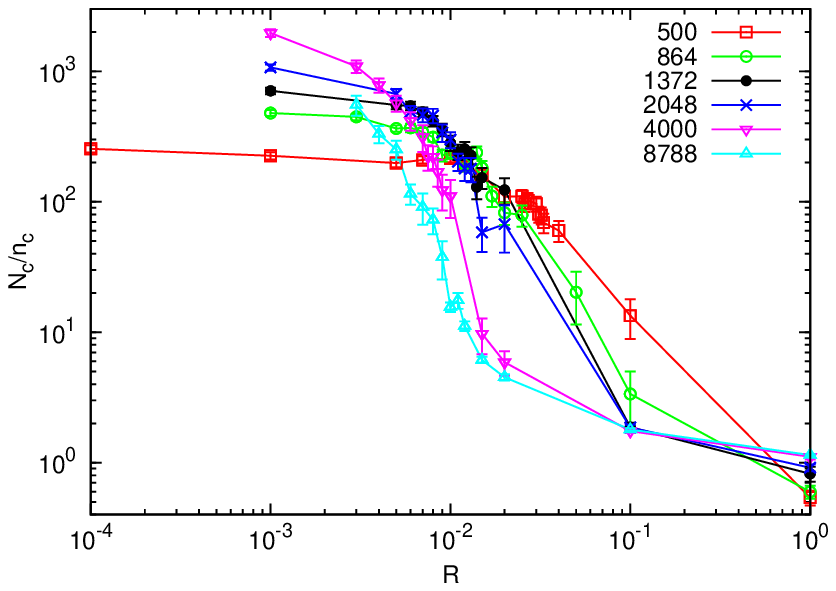}
\caption{(color online)  (left) The number of crystalline clusters $n_c$ (FCC, HCP,
and BCC) normalized by the system size $N$ for monodisperse LJ systems
as a function of cooling rate $R$ for several system sizes. (right)
The number of (FCC, HCP, and BCC) crystal-like particles $N_c$
normalized by the number of crystalline clusters $n_c$ ({\it i.e.}
average crystalline cluster size) as a function of cooling rate for
several system sizes.}
\label{fig:dxacluster}
\end{figure*}

We also employed a crystal analysis algorithm to
identify the crystalline clusters (FCC, HCP~\cite{stukowski:2012}, or
BCC) that form during the thermal quenching process.  
For slow cooling rates, the system forms only a few large crystalline
clusters whose size scales with the system size.
(See Fig.~\ref{fig:dxacluster}).  For fast cooling rates, the number of
crystalline clusters is small, and each cluster contains only a few
particles. At intermediate rates, the number of crystalline clusters
reaches a maximum at a characteristic cooling rate that scales with
$N$ as shown in Fig.~\ref{fig:dxacluster}. These results are
consistent with the fact that the critical cooling rate $R_c$ (defined
using the local bond orientational order parameter $Q_6^l$) becomes
independent of the system size in the $N\rightarrow \infty$ limit.

\end{appendix}


\begin{thebibliography}{29}
\bibitem[{foo()}]{footnote:1}
\bibinfo{note}{HCP-like particles and clusters reported in this work are
  identified by considering the first nearest neighbors of each particle. If
  instead, the super-lattice across stacking layers was included, most of the
  HCP clusters would be classified as `9R' structures~\cite{9r} (with repeating
  motifs composed of an FCC layer followed by two HCP layers).}
\end{thebibliography}

\end{document}